# Robust Material Properties in Epitaxial $In_2Te_3$ Thin Films Across Varying Thicknesses

Maximilian Buchta, Felix Hoff*, Lucas Bothe, Niklas Penner, Christoph Ringkamp, Thomas Schmidt, Timo Veslin, Ka Lei Mak, Jonathan Frank, Dasol Kim and Matthias Wuttig

M. Buchta, L. Bothe, K. L. Mak, Prof. M. Wuttig

    Peter-Grünberg-Institute – JARA-Institute Energy Efficient Information Technology (PGI-10) Wilhelm-Johnen-Straße, 52428 Jülich, Germany

F. Hoff, N. Penner, C. Ringkamp, T. Schmidt, T. Veslin, J. Frank, D. Kim, Prof. M. Wuttig

    Institute of Physics IA, RWTH Aachen University, Sommerfeldstraße, 52074 Aachen, Germany

    E-Mail: hoff@physik.rwth-aachen.de

This preprint was submitted for peer review at Small.

**Abstract**

Sesqui-chalcogenides serve as a critical bridge between traditional semiconductors and quantum materials, offering significant potential in applications such as thermoelectrics, phase change memory, and topological insulators. While considerable attention has been focused on antimony- and bismuth-based compounds - characterized by substantial property changes upon reduction in film thickness – indium containing sesqui-chalcogenides like $In_2Te_3$ are emerging as promising candidates for photovoltaics and electronic devices. However, the effects of film thickness on the properties of $In_2Te_3$ remain largely unexplored.

In this study, we investigate high-quality $In_2Te_3$ thin films grown by molecular beam epitaxy on Si (111) substrates across a thickness range from 2.7 nm to 24 nm. We employ X-ray diffraction, reflective high-energy electron diffraction and atomic force microscopy to analyze both the crystal structure and film morphology. Additionally, we utilize broadband optical spectroscopy alongside femtosecond pump-probe measurements and Raman spectroscopy to assess optical and vibrational properties, respectively.

Our analysis reveals that material properties exhibit minimal dependence on film thickness - contrasting sharply with behavior observed in other chalcogenides such as $Sb_2Te_3$, $Bi_2Se_3$, or GeTe. This phenomenon can be attributed to covalent bonding present in $In_2Te_3$, which differs from those in its antimony- and bismuth-containing counterparts.

**Introduction**

The family of sesqui-chalcogenides plays a pivotal role in connecting conventional semiconductors and emerging quantum materials, exhibiting applications as thermoelectric materials [1, 2], phase change memory materials [3, 4] and topological insulators [5, 6]. While significant attention has been directed towards antimony and bismuth-based sesqui-chalcogenides, indium-containing compounds have recently gathered considerable interest within the field of photovoltaics [7], ferroelectric data storage [8, 9] and electronic devices [10, 11]. Notably, $In_2Te_3$ has emerged as a promising material for broad-spectrum photodetectors [12, 13] and photovoltaic power generation [7, 14].

Despite its technological relevance, research into the fundamental understanding of thin film crystal structures and phase transitions remains in its early stages. The existing studies predominantly focus on bulk crystals [15-17], even though high-quality thin films are crucial for modern compact devices. Recently, research has shifted toward growing thin films using various deposition techniques across a range from a few to hundreds of nanometers (nm) [11, 12, 18].

As demonstrated for certain chalcogenides, material engineering based on film thickness is particularly effective when the thickness is reduced below 10 nm [19-23]. This phenomenon has been linked to a newly proposed type of chemical bonding termed metavalent bonding, which involves the interplay between electron localization and delocalization [19-24]. Such bonding results in high Born effective charges ($Z^*$), elevated dielectric constants, near-metallic conductivity, large mode-specific Grüneisen parameters for optical phonons, unusual bond-breaking behaviors, and effective coordination numbers (ECoN) that deviate from the octet rule [25-27]. These phenomena arise from the competition between electron localization and delocalization, not found in covalent crystals [28, 29].

Therefore, this study aims to investigate whether this phenomenon applies to In$_2$Te$_3$ as well. To achieve this goal, high-quality thin films with thicknesses ranging from 2.7 nm to 24 nm were grown on Si (111) using molecular beam epitaxy (MBE). The structural properties of these samples were analyzed concerning their thickness using low-energy electron diffraction (LEED), reflection high-energy electron diffraction (RHEED), X-ray diffraction (XRD), X-ray reflectometry (XRR), and atomic force microscopy (AFM). Their thickness dependent properties were investigated using broadband optical spectroscopy to obtain the dielectric function. Additionally, femtosecond pump-probe measurements combined with Raman spectroscopy were employed to characterize the lattice dynamics and light-matter interactions. Finally, density functional theory (DFT) is used to gain insights into the chemical bonding mechanism within the material. This comprehensive approach ensures a unified interpretation of the experimental findings.

1. **Epitaxially Grown In$_2$Te$_3$ Thin Films and their Atomic Structure**

Indium Telluride crystallizes in multiple stable configurations, including, among others, InTe, In$_3$Te$_4$, In$_2$Te$_5$, and In$_4$Te$_3$ [30]. Therefore, MBE is employed due to its ability to provide excellent stoichiometric control. The films were grown on the unreconstructed Si (111) 1 × 1 surface under Te-rich conditions. RHEED was utilized for in situ monitoring of the growth process. Subsequently, the samples were capped in-situ with Al$_2$O$_3$ using sputtering. Further details can be found in the Experimental section.

In$_2$Te$_3$ adopts the ZnS crystal structure [31], belonging to space group $F\bar{4}3m$, with a lattice constant of 6.16 Å (see Figure S1). To compensate for the additional electron contributed by indium compared to zinc, one third of the indium sites is vacant. The films are oriented along the (111) orientation of this crystal structure, resulting in an in-plane lattice constant of $a$ = 4.36 Å and an out-of-plane lattice constant of $c$ = 10.67 Å. The crystal structure of In$_2$Te$_3$ along the (111) growth direction is illustrated in **Figure 1**a for the in-plane direction and Figure 1b for the out-of-plane direction.

Figure 1c presents the LEED pattern of $In_2Te_3$ films, revealing two distinct in-plane rotational domains marked in blue and orange. A ring connecting these diffraction spots indicates an angular misalignment relative to the substrate – a phenomenon known as fiber texture. Yet, two rotational domains are predominantly found, which are also observed in the RHEED pattern oriented along the (11) in-plane direction of the substrate, as shown in Figure 1d. Shortly after growth commenced, three distinct film streaks appeared corresponding to these two rotational domains observed in LEED. Due to a 30 ° rotation between domains, they are assigned to the (10) order for domain 1 (blue), (11) order of domain 2 (orange) and (20) order of domain 1.

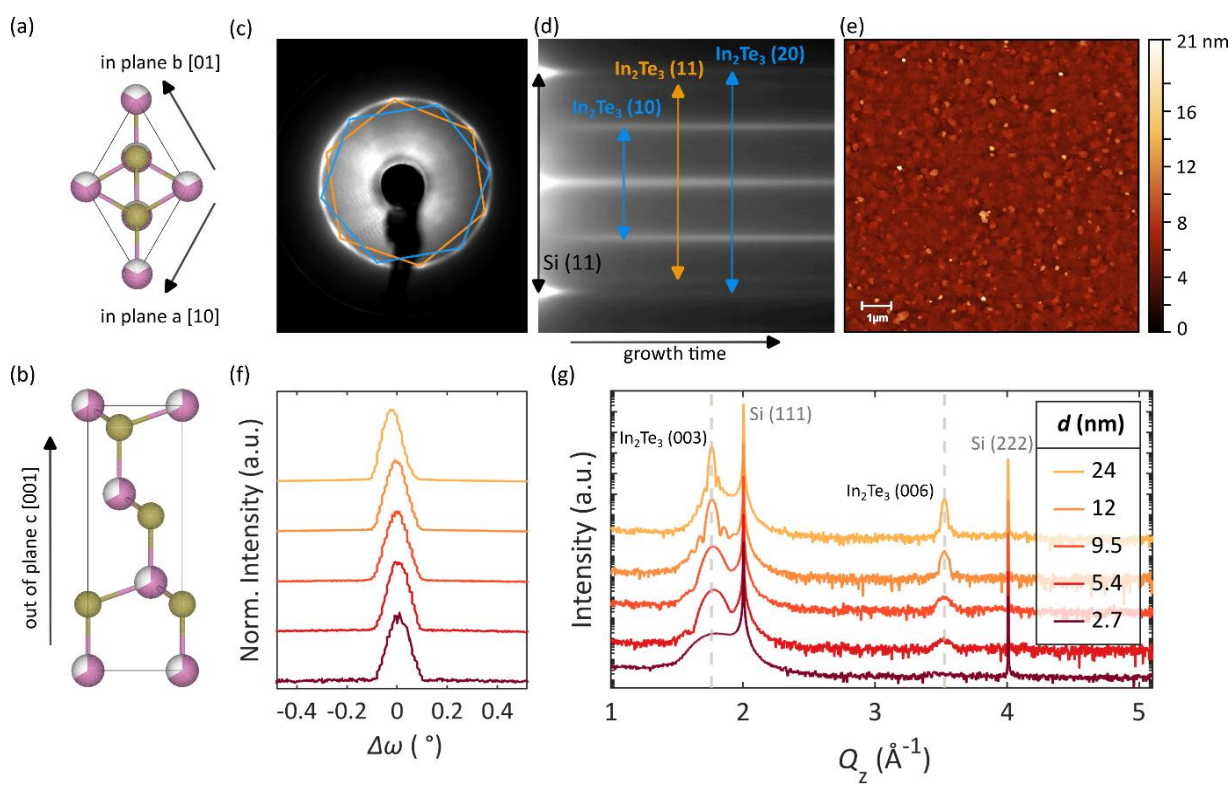

**Figure 1**: Structural analysis of the $In_2Te_3$ thin film series. The crystal structure of $In_2Te_3$ (111) is illustrated in both directions: (**a**) in-plane and (**b**) out-of-plane with the In atoms in pink and the Te atoms in gold. Panel **c)** displays the LEED pattern pattern (E = 59 eV) where the two rotational domains are marked in blue and orange; (**d**) shows the corresponding streaks from the RHEED growth video. Film streaks are labeled according to their respective in-plane diffraction order. The AFM scan in **e)** confirms a high surface quality with an RMS roughness value of 1.05 nm. The rocking curves and XRD spectra are displayed in **f)** and **g)**, that confirm a high texture quality alongside phase purity.

AFM has been employed to assess the surface morphology exemplified in Figure 1e for a sample without capping and a thickness of 9 nm, which exhibits a root mean square roughness (RMS) of 1.05 nm. This value aligns well with roughness measurements obtained via XRR, confirming consistency across methods used for thickness and roughness determination. Figure 1f presents rocking curves from the thin film samples exhibiting distinct peaks with full width half maximum (FWHM) values around approximately 0.1°. This observation corroborates the excellent texture within these thin films while indicating uniformity across different thicknesses within this series. Notably, peak positions shift slightly towards lower values as film thickness decreases.

Figure 1g presents the $\theta-2\theta$ diffractograms that confirm the high sample quality, as evidenced by the existence of distinct Laue fringes. These fringes arise from the interference of X-rays reflected from the top and bottom interfaces of a crystalline film with high structural order, i.e., smooth surfaces and interfaces as well as a uniform crystallinity, i.e., a single, well-oriented grain in the direction of the film normal. Specifically, clear oscillating patterns emerge for thicker samples' (003) peaks. Apart from the substrate and those corresponding to (003) and (006) peaks, no additional peaks are observed. The intensity of the (006) peak is less than that of the (003) peak, in line with the calculated powder diffraction pattern.

Pseudo-Voigt fits for both peaks have been used to determine the out-of-plane lattice constant $c$, using the average of the (003) and (006) results. For the thinnest sample, interference occurs between the (003) peak and the Si (111) peak. To address this, the silicon peak has been incorporated in the fit. For all remaining samples, contributions from silicon were not considered. The out-of-plane lattice constant of the thickest sample was determined to be $c$ = 10.69 Å. The in-plane lattice constant of the thickest sample was determined to be $a$

= 4.38 Å based on RHEED data. Both experimental in-plane and out-of-plane lattice constants show good agreement with literature values, with an increase of 0.2 % and 0.5 %, respectively.

**Figure 2** illustrates how both lattice constants change as the film thickness decreases. The in-plane lattice constant exhibits a slight increase of about 0.02 Å (0.4 %) between the thickness of 24 nm and 2.7 nm, as shown in the lower part of Figure 2. During initial growth stages, strong overlap occurred between (11) and (20) RHEED streaks with those from the substrate; therefore, only (10) streaks were used to extract the in-plane lattice constant.

Given the significant lattice mismatch between the Si (111) substrate and the $In_2Te_3$ film (3.86 Å versus 4.38 Å), which amounts to 11.9 %, it is surprising to observe an increase in the in-plane lattice constant. Instead, one would expect it to decrease and converge towards 3.86 Å. However, when considering the film's in-plane domain rotated by 30 ° relative to Si (111), a coincidence lattice can be identified, as shown in the supplementary information (Figure S2). This adjustment reduces the mismatch to 1.8 % and explains the observed increase in the in-plane lattice constant with decreasing film thickness as a result of strain from the substrate.

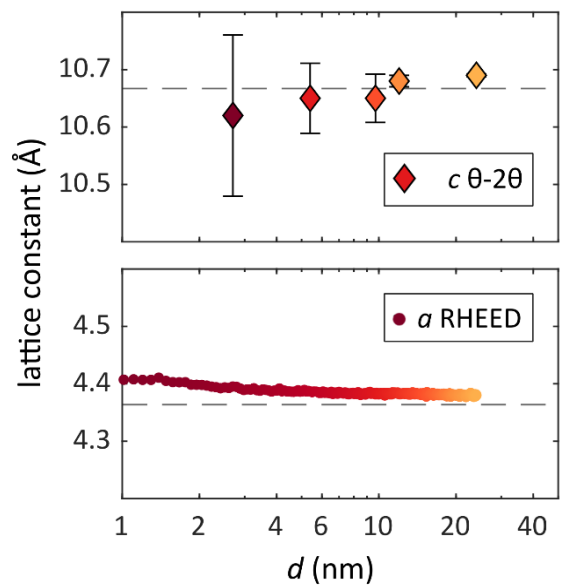

**Figure 2**: Thickness dependent lattice constants of $In_2Te_3$. The in-plane lattice constant has been extracted from the (10) streaks in the RHEED growth video, which displays an increase of 0.4 % between the thickness of 24 nm and 2.7 nm. The out-of-plane lattice constant has been extracted from the XRD spectra based on the average of (003) and (006) peaks. The literature bulk values shown as dashed lines [31].

On the other hand, the out-of-plane lattice constant *c* shows a slight decrease with decreasing film thickness by approximately 0.5 % between the thickest and thinnest films. This decrease correlates with the in-plane increase, that effectively preservers the unit cell volume. Notably, both lattice parameters converge towards the bulk literature values for film thicknesses above 12 nm.

## 2. Dielectric Function of In$_2$Te$_3$ Thin Films and its Response to Thickness Reduction

The consistent high structural quality throughout the range of film thicknesses is advantageous for the subsequent assessment of optical properties. To investigate the dielectric properties as a function of film thickness, the reflectance of the sample is recorded using Fourier-Transform Infrared (FTIR) and fiber grating spectroscopy over a wide spectral range from far infrared to near ultraviolet. The measured reflectance data are shown in Figure S3. This data is complemented by ellipsometry measurements from near infrared to ultraviolet. The dielectric function of In$_2$Te$_3$ was obtained by building a layer stack model and fitting the reflectance and ellipsometry data by using Tauc-Lorentz oscillators. The resulting dielectric functions for different thicknesses are displayed in **Figure 3**, the real part is shown in Figure 3a and the imaginary part in Figure 3b.

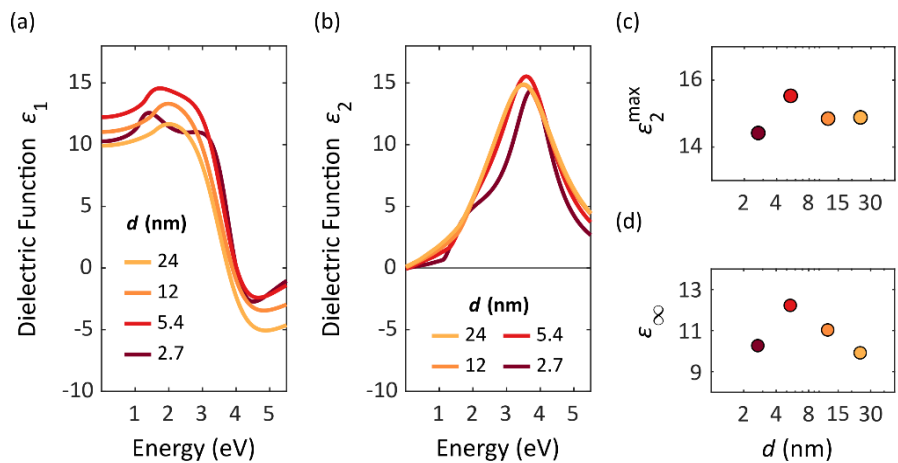

**Figure 3**: The real part a) and imaginary part b) of the dielectric function of four different In$_2$Te$_3$ thin films obtained by optical spectroscopy. (c) Maximum value of $\varepsilon_2(\omega)$ and the optical dielectric constant $\varepsilon_\infty$ (d) plotted as a function of film thickness.

The optical dielectric constant $\varepsilon_\infty$ has moderate values ranging from 10 to 13 and does not exhibit a clear trend with respect to film thickness, as shown in Figure 3d. This optical dielectric constant $\varepsilon_\infty$ is defined at frequencies above the highest phonon resonance but below the next natural frequency of the crystal, i.e., the smallest possible interband transition (band gap) [32]. To meet these criteria set forth in the definition, $\varepsilon_\infty$ values may be extracted at an energy of 0.05 eV.

The imaginary part of the dielectric function shows a distinct maximum associated with an interband transition at around 3.5 eV. Importantly, no significant dependence on film thickness is observed. For all samples, $\varepsilon_2^{max}$ is between 14 and 16 as shown in Figure 3c. Notably, the energy associated with the interband transition experiences a small shift of 0.3 eV towards higher energies upon reducing the layer thickness, transitioning from 3.5 eV to 3.8 eV, indicating a subtle change in absorption characteristics. To delve deeper into these phenomena, we will explore the thickness-dependent light-matter interactions and their implications for lattice dynamics in In$_2$Te$_3$ thin films in the subsequent section.

## 3. Lattice Dynamics of In$_2$Te$_3$ Thin Films and its Response to Thickness Reduction

The previous sections have demonstrated that the atomic structure and optical properties depend only slightly on film thickness. Consequently, we expect that photoexcitation and the resulting lattice dynamics will also remain largely unchanged with varying film thicknesses. To investigate the dynamics of the charge carriers and the crystal lattice in response to an external stimulus, femtosecond pump-probe measurements were conducted.

**Figure 4**a displays the isotropic transient reflectance for films of different thicknesses, with curves vertically shifted for clarity. The initial change in reflectance observed within the first few hundred femtoseconds is due to the direct response of the excited carriers and the formation of non-equilibrium excited states [33]. This excess energy is partially converted into coherent optical phonons, which can be observed for all film thicknesses [33-35]. To determine the frequencies of these coherent phonons, we first subtract the background using a phenomenological model that describes the initial excitation of carriers and their subsequent relaxation. The oscillatory part of the signals can be Fourier transformed to extract phonon frequencies, illustrated in Figure 4b.

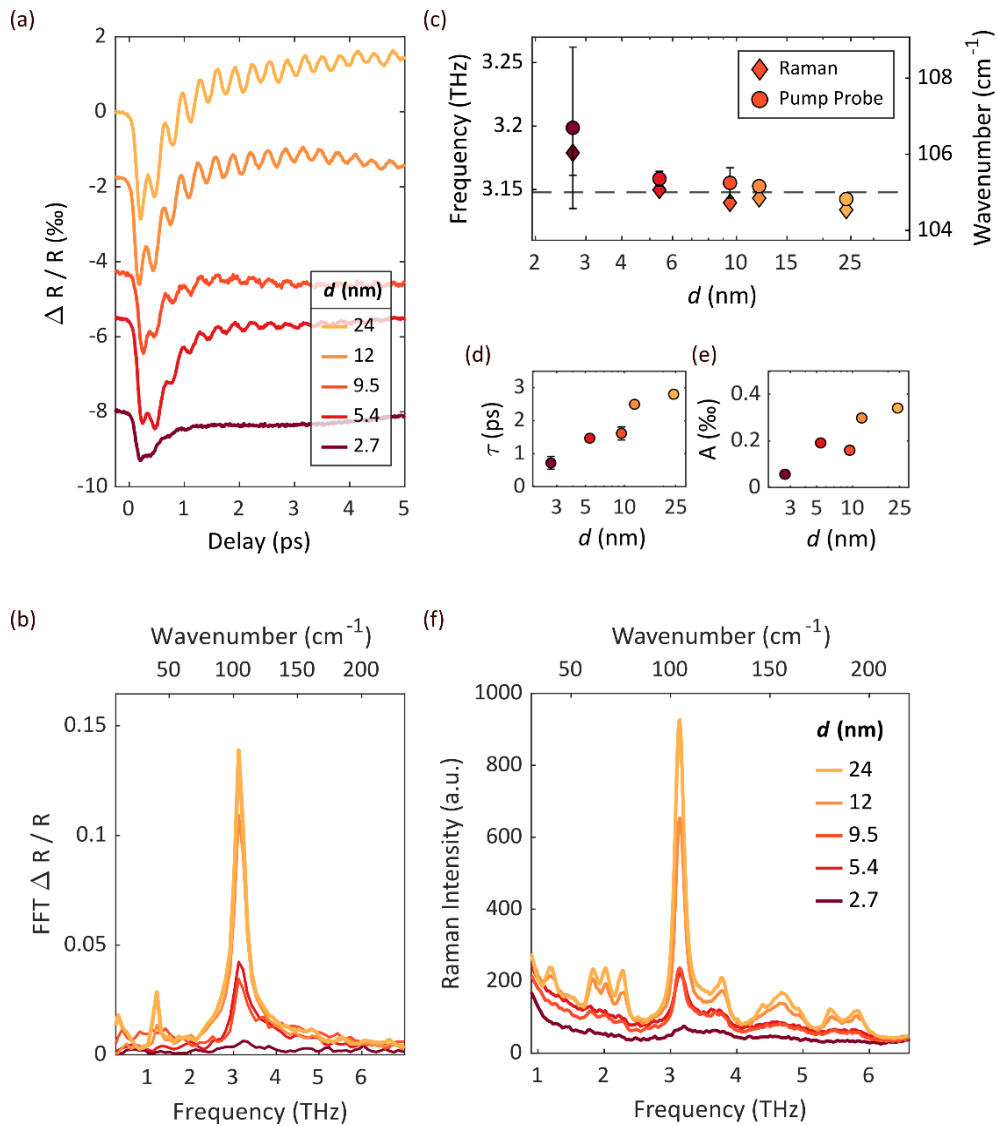

**Figure 4:** (**a**) Isotropic transient reflectance for the thin film samples. Curves are vertically shifted for clarity. The excitation of coherent phonons is visible in each curve. (**b**) Fourier transformation of the oscillatory component of the signal showing two distinct phonon frequencies. The mode frequencies resulting from damped harmonic oscillator fits are shown in (**c**). The corresponding amplitudes and decoherence times against their film thickness are in (**d, e**). (**f**) Raman Shift data for different film thicknesses. Peak positions from Raman spectroscopy data are shown as diamonds in (c).

A dominant peak at about 3.15 THz (105 cm$^{-1}$) is observed for all film thicknesses, while a smaller peak at 1.2 THz (40 cm$^{-1}$) is present only in the two thickest films. These modes align well with the Raman literature values at 105 cm$^{-1}$ and 42 cm$^{-1}$ [36, 37], where the phonon mode at 105 cm$^{-1}$ has been identified as an $A_{1g}$ symmetry phonon [37]. This observation corresponds well with its dominance in isotropic transient reflectance measurements, which primarily resolve fully symmetric $A_{1g}$ phonons [35].

A more accurate understanding is achieved by fitting these oscillations with damped harmonic oscillators. The resulting frequencies show a slight increase - approximately 1 % - with decreasing film thickness, as depicted in Figure 4c. Both decoherence time $\tau$ and amplitude $A$ of the reflectance changes generated by coherent phonons decrease as film thickness diminishes. This trend is illustrated in Figures 4d and e. Specifically, decoherence time decreases from 3 ps to below 1 ps, while amplitude drops from 0.4 ‰ to 0.08 ‰.

Additionally, Raman spectroscopy provides insights into vibrational properties without optical pre-excitation. The corresponding spectra are presented in Figure 4f. The phonon peak at 105 cm$^{-1}$ dominates the spectrum across all layer thicknesses; however, the previously noted peak at 40 cm$^{-1}$ was resolved only for the two thickest samples. By employing Lorentzian fits, we determined peak positions, which are compared to the pump-probe data shown in Figure 4c, exhibiting good agreement. It should be noted that we identified a total of 11 distinct phonon peaks in the data from the thickest sample which align well with existing literature on bulk samples [36]. This is further discussed in the Supporting Information (Figure S4).

## 4. Discussion

The results presented demonstrate that thin In$_2$Te$_3$ films grown on Si (111) exhibit only minor dependency on thickness. Specifically, a relative change of below 1 % is observed in both lattice parameters. Furthermore, no substantial shifts in the dielectric constant or the height of the absorption maximum are reported. The frequency of the dominant $A_{1g}$ phonon mode increased moderately by about 1 %.

This behavior contrasts sharply with that of other sesqui-chalcogenides, monochalcogenides, and pnictogens [21, 22]. For instance, $Sb_2Te_3$, GeTe, and Bi show a relative change exceeding 5 % in the frequency of their $A_{1g}$ phonon modes with comparable reductions in film thickness [19, 20, 22]. In these materials, the out-of-plane lattice constant *c* shows a pronounced dependency on sample thickness and increases for thinner films, contrary to the results presented here. Additionally, the dielectric functions of these materials change significantly, both the dielectric constant and the height of the absorption maximum are reduced by half. Concurrently, the position of the absorption maximum shifts towards higher photon energies [21].

The distinct behavior of $In_2Te_3$ can be closely linked to its chemical bonding mechanisms, different to other sesqui-chalcogenides. Several bond indicators support this observation: a low Effective Coordination Number (ECoN) of 4; a low mode-specific Grüneisen parameter ranging from 1.5-2.1 [38]; and a room temperature conductivity measured at around $10^{-1}$ S/cm [38], all suggesting covalent bonding in $In_2Te_3$. The low values shown for its dielectric function in Figure 3 further support this characterization as they align with those found in other covalently bonded semiconductors.

Two quantum-chemical bond descriptors, electrons shared (ES) and electrons transferred (ET), are introduced to characterize electron distribution while varying the arrangement of vacancies [39, 40]. As shown in Figure S5, the ES values of all structures range from 1.6 to 2.2, accompanied by moderate ET values (< 0.25). These bonding descriptors fall within the range characteristic of covalent bonding, regardless of the specific vacancy distribution [39, 40].

In contrast, Sb$_2$Te$_3$, GeTe, and Bi exhibit a different bonding mechanism characterized by a competition between electron localization and delocalization due to a network of overlapping *p*-orbitals. This results in a unique conductivity range (between 10$^3$ and 10$^4$ S/cm) that situates these materials between semiconductors and metals, earning them the designation 'incipient metals.' [25] These incipient metals share a common characteristic: their chemical bonding between atoms is primarily influenced by half-filled *p*-orbitals [24, 25]. In reduced dimensions, the competition between electron localization and delocalization is modified and an increased degree of electron localization alters material properties and crystal structures [21, 23].

For metavalent materials, out-of-plane relaxation occurs through a mechanism where electron localization and delocalization adjust upon changing film thickness to minimize energy by altering atomic positions [21, 22]. This mechanism has been documented extensively for thin GeTe layers [41]. Importantly, the metavalently bonded thin films expand their out-of-plane lattice constant by several percent upon reduction in thickness, in In$_2$Te$_3$ it reduces.

Structural distortions within thin films due to variations in bond strengths significantly influence phonon mode frequencies. For layered chalcogenides with n-tuple stacking along the (001) direction - specifically Sb$_2$Te$_3$ - the $A_{1g}^1$ mode exhibits a redshift while the $A_{1g}^2$ mode shows a blueshift. Consequently, no strong trend emerges regarding phonon frequencies in In$_2$Te$_3$ (as shown in Figure 4), since significant changes do not occur within chemical bonding and thus atomic structure when layer thickness is reduced.

**5. Conclusion**

In summary, this study has examined the impact of thickness on the properties of thin In$_2$Te$_3$ films. The results demonstrate that the intrinsic material properties of covalent In$_2$Te$_3$ films remain nearly constant even when film thickness is drastically reduced. This behavior contrasts sharply with that of metavalent sesqui-chalcogenides such as $Sb_2Te_3$, $Bi_2Te_3$ or $Bi_2Se_3$, which exhibit significant changes in bonding and intrinsic properties upon reduction in film thickness. Only small changes were observed in the structural, optical, and vibrational properties of $In_2Te_3$ films. This finding suggests that other covalent sesqui-chalcogenides such as $Bi_2S_3$, $Sb_2Se_3$ and $Sb_2S_3$ may also experience only modest changes in their intrinsic material properties when film thickness is reduced. Despite differences in crystal structure among these materials, they share similar chemical bonding characteristics with $In_2Te_3$.

## 6. Methods

*Growth:* Prior to insertion into the UHV chamber, the substrates (Okmentic, Si(111), SSP, p doped, > 5 kΩ) underwent a cleaning process with H2SO4:H2O2. Subsequently, they were immersed in HF to remove the native oxide layer and passivate the dangling bonds with hydrogen.

The substrates were heated to T = 330 °C in UHV to evaporate the hydrogen and subsequently cooled down to the growth temperature of T = 210 °C. Note that all temperature readings have an offset due to the thermocouple reading. The unreconstructed surface was validated using RHEED. Optimal growth conditions for growth along the [111] orientation was determined in a tellurium-rich environment with a beam flux ratio In:Te of 1:20. Both elements were evaporated from standard effusion cells. The growth rate was 0.1 nm/min, monitored by RHEED. After growth, the samples were capped in-situ with an $Al_2O_3$ layer with a thickness of approximately 10 nm, employing sputtering.

*LEED:* After growth, the sample has been *in*-situ transferred to the LEED chamber, where the LEED pattern was acquired with a Specs ErLEED 1000-A Optics and a background pressure of $1*10^{-10}$ mbar.

*AFM*: AFM measurements have been performed for a sample without capping using a Burker Dimension Edge AFM. The scan size was 10 x 10 µm and 512 x 512 scans were taken with a tapping frequency of 0.5 Hz.

*X-Ray Reflection and Diffraction:* The measurements were measured on a "Rigaku Smartlab" system with a rotating anode employing a Cu Kα1 radiation source (λ = 1.54 Å), selected by a Ge (220) 2 bounce monochromator.

*Femtosecond Pump Probe Reflectivity:* Ultrafast optical measurements were conducted using a standard reflection-type, two-color pump-probe experiment in an isotropic configuration. The 800 nm wavelength, 60 fs width pump beam was separated from a Ti:Sapphire regenerative femtosecond amplifier, chopped at 1500 Hz, and directed to a free-standing optical delay line before being focused on the sample to a spot size of 200 µm in diameter.

The probe pulses were frequency converted to 516 nm via sum frequency generation in an optical parametric amplifier and focused to a 30 µm diameter spot on the sample. The detection unit comprised two balanced Si photodiodes connected to variable gain current amplifiers and a data acquisition card. All measurements were taken at incident pump fluences of around 1 mJ/cm², with the probe fluence being ten times lower. To eliminate systematic errors on laboratory time scales, the order of data point recording and the positioning of the delay line were randomized. The isotropic transient reflectance was calculated from the whole signal: $\Delta R = \frac{\Delta(R_\text{s}+R_\text{p})}{R_0}$. A polarizing beam splitter cube was used to split the reflected probe beam correspondingly. Transient reflectance was normalized to the steady state reflectance $R_0$. Reversibility of optical excitation was ensured by monitoring the static reflectivity gained for the probe-pulse when the pump-pulse was chopped.

*Raman Spectroscopy:* Raman spectroscopy measurements were conducted with a commercial WITec system (alpha300), operating in backscattering geometry. A diode pumped solid state laser supplied linearly polarized light with a wavelength of 531 nm (2.33 eV). This was delivered through a single-mode optical fiber and used to excite the sample. The spot size of 400 - 500 nm was obtained by employing a long working focusing lens with a numerical aperture of 0.70. By setting the excitation power to 500 µW, significant heating effects were avoided. For detection, the reflected light was delivered through a single-mode optical fiber to a charge coupled spectrometer with a grating of 2400 lines per mm. All measurements were obtained at room temperature with a 50× (NA = 0.7) objective, with a resolution close to the diffraction limit (≈ 500 nm).

*Optical Spectroscopy:* Ellipsometry spectra were measured using three angles of incidence (65°, 70° and 75°) on a J.A.Woollam M-2000UI spectroscopic ellipsometer. The deuterium and halogen lamps served as the sources of illumination for the setup. A silicon CCD camera detected visible and ultraviolet light, while an InGaAs diode array captured lower energy photons. In total, 584 channels with an average of 7 meV resolution over 0.72 to 5.14 eV were available. Normal incidence reflectance data were collected by fiber grating in the range from 9100 to 45 000 cm$^{-1}$ with an Avantes AvaSpec-ULS2048CL-EVO with the use of a CMOS linear image sensor. Deuterium and halogen lamps (AvaLight-DH-S-BAL) were used as illumination sources. An aluminum mirror served as reference. Reflectance measurements in the 200 respectively 100 to 18 000 cm$^{-1}$ range were performed on a Bruker Vertex 80v Fourier-transform infrared spectrometer (FTIR). A 200 nm thick silver mirror served as a reference. Multiple source (Hg-arc, globar, and tungsten lamp), beam splitter (Mylar 50 µm, multilayer, KBr, and CaF2), and detector (DLaTGS (one with polyethylene and one with KBr window), liquid nitrogen-cooled InSb and room-temperature Si photodiode) combinations were utilized to cover the entire spectral range. The dielectric functions of the samples were determined using a five-layer model within the J.A. Woollam CompleteEASE 6 software. The bottom layer consists of the silicon substrate, followed by the indium telluride thin film. The dielectric function of the thin film was described by employing a summation of two Tauc-Lorentz oscillators without a Drude contribution. The top two layers represent the capping in an effective medium approach (EMA) to account for a mixture of aluminum oxide with and without a slight amount iron impurity, both of which were measured independently in advance. An EMA layer describing a roughness layer obtained from X-ray reflectivity (XRR) measurements is situated between the capping layers and the thin film. All EMA layers were fitted using the Bruggeman model, and all parameters were adjusted until convergence was achieved.

*DFT Studies:* To validate our experimental observations, density functional theory (DFT) are carried out using the Vienna Ab Initio Simulation Package (VASP) with the projector-augmented wave pseudopotentials [42-45]. The Perdew–Burke–Ernzerhof functional is employed to approximate the exchange–correlation energies of electrons [46]. The cutoff energies for the plane-wave basis are set to 550 eV. The quantification of electrons shared (ES) within chemical bonding utilizes the density-derived electrostatic and chemical (DDEC) approach [47-49]. ES is quantified as twice the value of the bond order.

*Proofreading*: OpenAI GPT 4o was used to check grammar, syntax and clarity. The content and scientific accuracy of the paper remain the responsibility of the authors.

## Supporting Information

Supporting Information is available online or from the author.

## Author Contribution

M. Buchta and F. Hoff contributed equally to the project. M. Buchta initiated the project. M. Wuttig supervised the project. M. Buchta, C. Ringkamp K.L Mak and L. Bothe fabricated the samples and performed XRD, XRR, RHEED, LEED and AFM measurements and analyzed them. T. Schmidt and N. Penner conducted optical spectroscopy measurements and data analysis. T. Veslin, J. Frank and F. Hoff performed pump probe and Raman measurements and analyzed the data. D. Kim calculated ES values. F. Hoff prepared the figures. M. Buchta and F. Hoff wrote the manuscript. M. Wuttig significantly revised the manuscript. All authors commented on and approved the submission of this manuscript.

## Acknowledgments

F. Hoff, T. Veslin, T. Schmidt and M. Wuttig acknowledge financial support from NeuroSys as part of the initiative "Clusters4Future", which is funded by the Federal Ministry of Education and Research BMBF (03ZU2106BA). M. Buchta, L. Bothe and M. Wuttig acknowledge financial support from Federal Ministry of Education and Research (BMBF, Germany) in the project NEUROTEC II (16ME0398K). The authors would like to express their gratitude to the entire team at the Helmholtz Nano Facility and the Nanocluster for their outstanding work, with a special thanks to Benjamin Bennemann.

**Data Availability Statement**: The data that support the findings of this study are available from the corresponding author upon reasonable request.

**Conflict of Interest**: The authors declare no conflict of interest.

Maximilian Buchta, Felix Hoff*, Lucas Bothe, Niklas Penner, Christoph Ringkamp, Thomas Schmidt, Timo Veslin, Ka Lei Mak, Jonathan Frank, Dasol Kim and Matthias Wuttig

M. Buchta, L. Bothe, K. L. Mak, Prof. M. Wuttig

    Peter-Grünberg-Institute – JARA-Institute Energy Efficient Information Technology (PGI-10) Wilhelm-Johnen-Straße, 52428 Jülich, Germany

F. Hoff, N. Penner, C. Ringkamp, T. Schmidt, T. Veslin, J. Frank, D. Kim, Prof. M. Wuttig

    Institute of Physics IA, RWTH Aachen University, Sommerfeldstraße, 52074 Aachen, Germany

    E-Mail: hoff@physik.rwth-aachen.de

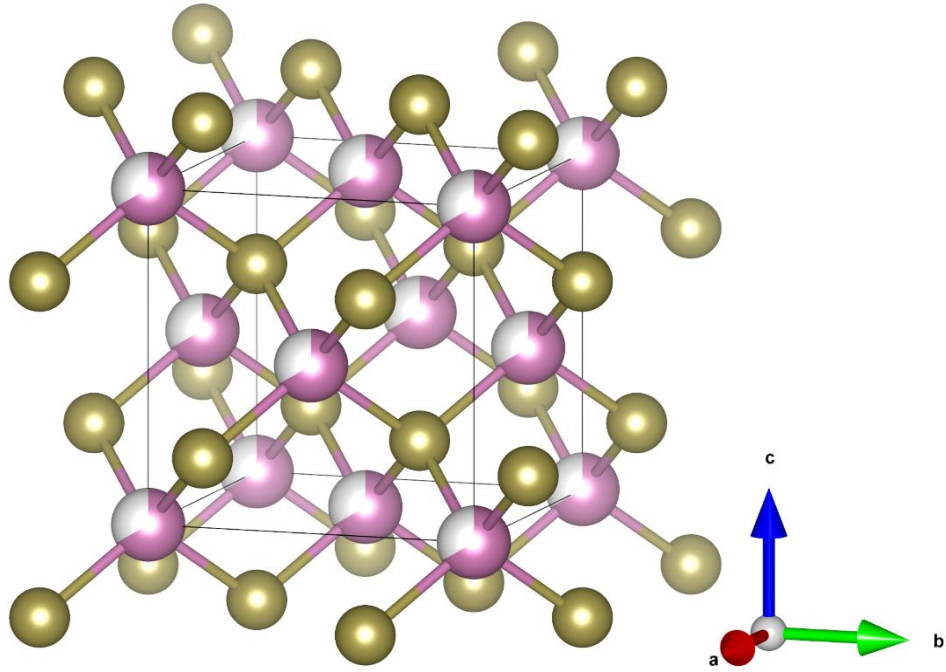

**Figure S1**: In$_2$Te$_3$ adopts the ZnS crystal structure [31], belonging to space group $F\bar{4}3m$. To compensate for the additional electron contributed by indium (displayed by the pink atoms) compared to zinc, the indium site is vacant by one third, homogenously distributed throughout the crystal.

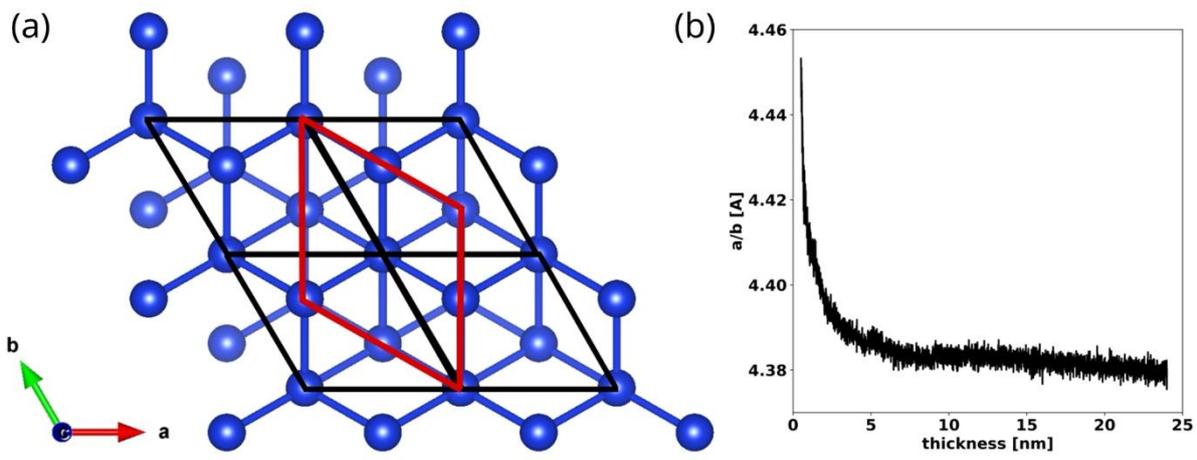

**Figure S2**: (**a**) Coincidence lattice of In$_2$Te$_3$(111) (red) and Si(111) (unit cell in black and atoms in blue) with a coincidence lattice constant of a = 4.46 Å, corresponding to a strain of 2.4 %. This alignment helps minimize strain at the interface between the substrate and the thin film. (**b**) RHEED data showing the evolution of the in-plane lattice constant during growth. Initially, the in-plane lattice constant is approximately 4.45 Å, indicating that the growth starts under strain towards the coincidence site lattice. This value sharply decreases within the first 2 nm of film thickness, after which the lattice constant gradually relaxes and converges towards the bulk value of In$_2$Te$_3$ as thickness increases.

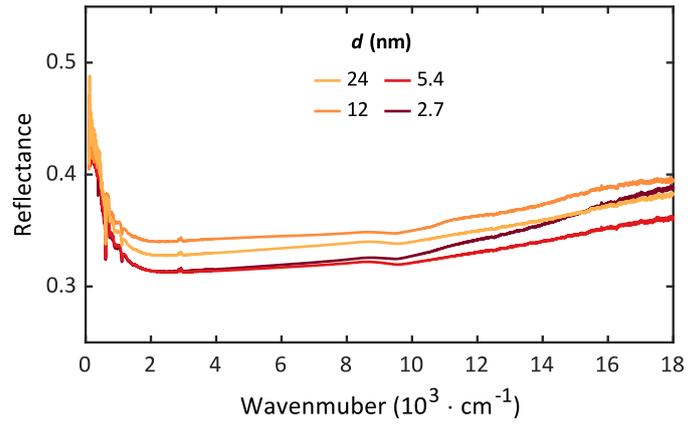

**Figure S3**: Reflectance raw data of the films of different thicknesses.

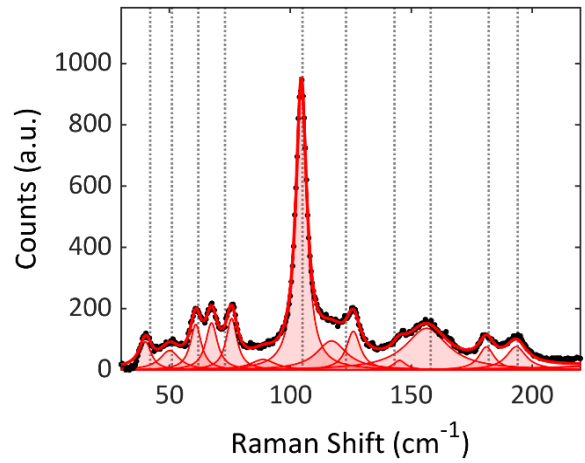

**Figure S4**: Raman data of the 24 nm sample with Lorentzian fits (red) and literature values from [36] (dotted lines).

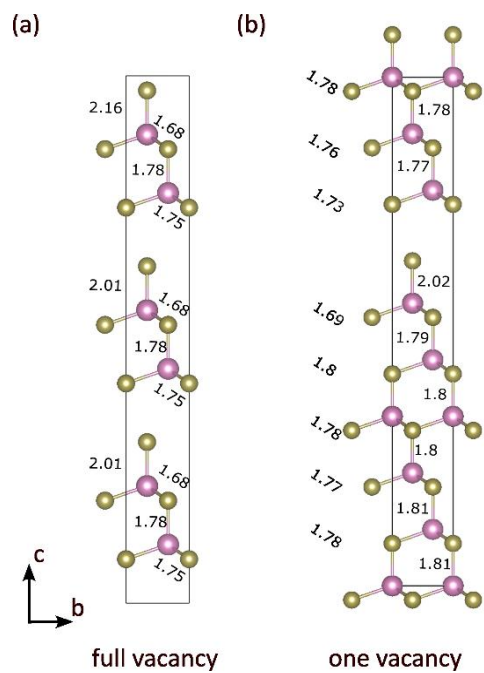

**Figure S5**: Electrons shared (ES) values for two distinct unit cells with different vacancy positions, as homogenous vacancy concentrations cannot be incorporated in these calculations. All ES values fall within a range of 1.6 to 2.2, confirming covalent bonding between the atoms in each unit cell.

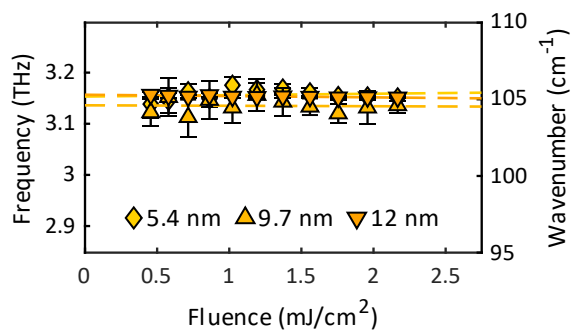

**Figure S6**: Coherent phonon frequencies remain stable across a broad range of incident pump fluences for three different film thicknesses (5 nm to 12 nm). Dashed lines show linear fits, highlighting the absence of phonon softening or hardening with increasing fluence.